\documentclass[twocolumn,showpacs,preprintnumbers,prb]{revtex4}

\usepackage{amsmath}
\usepackage{amssymb}
\usepackage{psfrag}
\usepackage{pstricks}
\usepackage{graphicx}
\usepackage{dcolumn}
\usepackage{bm}

\newcommand{\fig}[2]{\includegraphics[width=#1]{#2}}

\newcommand{\ud}{\mathrm{d}}
\newcommand{\vg}{V_{\!g}}
\newcommand{\vf}{v_{\mathrm{\scriptscriptstyle F}}}
\newcommand{\dx}{\!\!\ud x\,}

\begin{document}

\title{ Fabry-Perot interference and spin filtering in carbon
nanotubes}

\author{Claudia S.\ Pe\c ca}
\author{Leon Balents}%
\affiliation{
Physics Department,
  University of California, Santa Barbara, CA 93106
}%

\author{Kay J\"org Wiese}
\affiliation{Kavli Institute for Theoretical Physics, University of
  California, Santa Barbara, CA 93106
}%

\date{\today}

\begin{abstract}
  We study the two-terminal transport properties of a metallic
  single-walled carbon nanotube with good contacts to electrodes,
  which have recently been shown [W.\ Liang et al, Nature 441, 665-669
  (2001)] to conduct ballistically with weak backscattering occurring
  mainly at the two contacts.  The measured conductance, as a function
  of bias and gate voltages, shows an oscillating pattern of quantum
  interference.  We show how such patterns can be understood and
  calculated, taking into account Luttinger liquid effects resulting
  from strong Coulomb interactions in the nanotube.  We treat
  back-scattering in the contacts perturbatively and use the Keldysh
  formalism to treat non-equilibrium effects due to the non-zero bias
  voltage.  Going beyond current experiments, we include the effects
  of possible ferromagnetic polarization of the leads to describe spin
  transport in carbon nanotubes.  We thereby describe both incoherent
  spin injection and coherent resonant spin transport between the two
  leads.  Spin currents can be produced in both ways, but only the
  latter allow this spin current to be controlled using an external
  gate.  In all cases, the spin currents, charge currents, and
  magnetization of the nanotube exhibit components varying
  quasiperiodically with bias voltage, approximately as a
  superposition of periodic interference oscillations of spin- and
  charge-carrying ``quasiparticles'' in the nanotube, each with its
  own period.  The amplitude of the higher-period signal is largest in
  single-mode quantum wires, and is somewhat suppressed in metallic
  nanotubes due to their sub-band degeneracy.
\end{abstract}

\pacs{73.63.-b,71.10.Pm,72.25.-b}

\maketitle

\section{\label{intro} Introduction}

Spin transport represents a new branch in mesoscopic physics with
several technological
applications\cite{Aronov,Silsbee,Johnson,Prinz,Jay,Gijs},
e.g.\ information storage, magnetic sensors and potentially quantum
computation\cite{Bennet}.  While most theoretical models are based on
Fermi liquid theory, some work has been done on strongly correlated 1D
systems using Luttinger liquid
theory\cite{balegger1,balegger2,si1,si2,spinBena}.  This work has
focused in the weak tunneling regime between the ferromagnet and the
1D system and found that spin transport may provide experimental
evidence of spin charge separation, one of the main predictions of
Luttinger liquid theory that remains to be observed experimentally in
an unambiguously accepted way.  Despite the possible technological
applications and contributions to the study of spin charge separation
in strongly correlated systems, very little experimental work has been
carried out on spin transport in 1D systems.\cite{expFM} This work is
complicated by the use of multi-walled carbon nanotubes, and explored
only situations with ferromagnetic contacts with parallel or
antiparallel magnetizations. 

\begin{figure}[b]
\centerline{\fig{.25\textwidth}{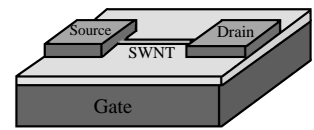}}
\caption{\label{scheme} Experimental geometry (from
Ref.~\onlinecite{Liang}). A single-walled carbon nanotube is located
on a silicon gate and oxide layer. The electrodes, which may be
ferromagnetic, are grown on top of the nanotube. The doped silicon is
used as a gate electrode to modulate the charge density. The
electronic transport properties of the nanotube devices were
characterized as a function of bias ($V$) and gate ($V_{\!g}$)
voltages.  }
\end{figure}

Early experimental work with nanotube devices was limited by poor
contacts between the electrodes and the nanotube, and accordingly
theoretical models focused in the weak tunneling regime.  Recently,
Liang {\textit{et al.}} \cite{Liang} have succeeded in fabricating
single-walled carbon nanotube devices with near-perfect ohmic contacts
to the electrodes. A schematic representation of their experiment is
presented in Fig.~\ref{scheme}. These devices are characterized by
values of the conductance as high as $G=3.7 e^2/h$, close to the
theoretically predicted higher limit\cite{datta} of $4 e^2/h$. At
temperatures below 10K, the measured conductance exhibits
approximately periodic oscillations as a function of the gate voltage. These
oscillations are due to Fabry-Perot interference -- i.e.\ quantum
interference between propagating electron waves inside the resonant
cavity defined by the two nanotube-electrode interfaces.
In order to explain their result, Ref.~\onlinecite{Liang} considered a
model of non-interacting electrons and used the multi-channel
Landauer-B\"uttiker formalism to calculate the differential
conductance as a function of the bias and gate voltages.  They have
found qualitative agreement between the calculated
conductance and their experimental data, especially with regard to the
variation of the low-bias conductance with gate voltage. 

On the other hand, transport experiments on carbon
nanotubes\cite{Bockrath,Yao1,Yao2,Dekker} have shown that electrons in
nanotubes are strongly correlated and are better described by a
Luttinger liquid
model\cite{Tomonaga,Luttinger,Haldane1,Haldane2,Egger,kbf}.  This
implies the electrons in these systems do not exhibit Fermi liquid
properties but instead form collective excitations better described by
charge and spin like density waves that propagate with different
velocities. This Luttinger liquid behavior drastically changes the
charge conductance for these systems and it is interesting to know how
this affects the results obtained for the particular setup used in
Ref.~\onlinecite{Liang}.
Furthermore, this setup can be generalized to the use of ferromagnetic
electrodes, in order to study both charge and spin transport in 1D
electron systems.

In this paper we investigate the spin and charge transport properties
in 1D electron systems with near-perfect contacts to ferromagnetic
electrodes (the normal metal electrodes correspond to the particular
case of zero magnetization). We consider both the case of quantum
wires, i.e.\ single-channel electron systems, and single-walled carbon
nanotubes, but mainly focus on the latter one. We use a
non-interacting Stoner model to treat the ferromagnetic leads and a
Luttinger liquid model for the nano\-tube and consider the case of
near-perfect contacts to the leads, therefore treating backscattering
at the contacts perturbatively. In order to introduce the effect of a
finite bias voltage, we use the non-equilibrium Keldysh formalism.
Following this procedure, we obtain the conductance, spin and spin
current as functions of the gate and bias voltages, the external
magnetic field and the orientation of the magnetization in each lead.
We study how the strong Coulomb interactions affect these transport
properties and find some features in the Fabry-Perot interference
patterns that are related to spin charge separation.

\section{\label{model}The model}

A single-walled carbon nanotube  with long-ran\-ge Cou\-lomb interactions
is well described by a forward-scatter\-ing model\cite{Egger,kbf}. In this
model the Hamiltonian density is given by 
\begin{eqnarray}\label{H}
{\mathcal H}_{\mathrm{LL}}&=&-i \vf \sum_{a=1}^2
\sum_{\alpha=\uparrow,\downarrow} \left( \psi_{Ra\alpha}^\dag\,
\partial_x \psi_{Ra\alpha}^{\vphantom{\dag}} - \psi_{La\alpha}^\dag
\,\partial_x \psi_{La\alpha}^{\vphantom{\dag}} \right) \nonumber \\
&+& \lambda \left[\sum_{a=1}^2 \sum_{\alpha=\uparrow,\downarrow}
\left( \psi_{Ra\alpha}^\dag\, \psi_{Ra\alpha}^{\vphantom{\dag}} +
\psi_{La\alpha}^\dag \, \psi_{La\alpha}^{\vphantom{\dag}}
\right)\right]^2  ,
\end{eqnarray}
where the right/left moving electron operators $\psi_{R/L\,a\alpha}$
have the labels $a=1,2$ for the band and $\alpha =\uparrow,
\downarrow$ for the spin projection of the electrons in the nanotube
and $\lambda$ is the interaction strength. The term in the square
brackets corresponds to the electron density.  We also consider
the same problem for a single-channel quantum wire, for which there is no
sub-band degeneracy and the band index can be dropped.  Due to the
similarities between the two cases, we  give explicit analytical
formulas throughout the paper only for the nanotube, but will present
results for the quantum wire where appropriate.

The metallic leads are modeled as two semi-infinite 1d-non-interacting
systems\cite{Safi,Maslov,Ponomarenko}, which is obtained with a position
dependent $\lambda$, constant in the wire and  zero in the leads.

\begin{figure}
\centerline{\fig{.3\textwidth}{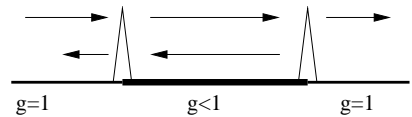}}
\caption{\label{fig:model} Schematic representation of the model.
\\ The leads are modeled as 1D non-interacting electron systems, the
Luttinger liquid parameter is therefore $g=1$. The nanotube (bold
line), on the other hand, is described by a 1D interacting system, in
this case $g<1$, which corresponds to repulsive interactions. The
contacts to the leads are modeled as two weak backscattering
barriers. The two backscatterers generate Fabry-Perot interference.}
\end{figure}

We allow for ferromagnetism in the leads using a non-interacting
Stoner model\cite{balegger1,balegger2} (mean-field treatment of the
magnetization).  The Hamiltonian density is ${\mathcal
H}_{\mathrm{FM}}={\mathcal H}_{0}+{\mathcal H}_{M}$ with ${\mathcal
H}_{0}={\mathcal H}_{LL} (\lambda =0)$ and
\begin{equation}
{\mathcal H}_{M} = - \vec{M} \cdot \sum_{a\alpha\beta}\left(
\psi_{Ra\alpha}^\dag \, \vec{\sigma}_{\alpha \beta}
\,\psi_{Ra\beta}^{\vphantom{\dag}}+ \psi_{La\alpha}^\dag\, \vec
\sigma_{\alpha \beta} \, \psi_{La\beta}^{\vphantom{\dag}} \right) \ ,
\end{equation}
where $\vec \sigma_{\alpha \beta}$ are the Pauli matrices and
$\vec{M}$ is the ``exchange field'', which is proportional
to the magnetization. This is constant in each ferromagnetic lead,
i.e.\ $\vec{M}\parallel\hat{m}_1$ for $x<-L/2$ and
$\vec{M}\parallel\hat{m}_2$ for $x>L/2$ and in ordinary paramagnetic
leads $\vec{M}=0$. In this case, the total system corresponding to a
nanotube between two ferromagnetic leads is described by the
Hamiltonian
\begin{equation}
H= \int_{|x|>L/2}\!\!\!\!\!\!\! \dx \ {\mathcal H}_{\mathrm{FM}} +
\int_{|x|<L/2} \!\!\!\!\!\!\! \dx\ {\mathcal H}_{\mathrm{LL}} \ .
\end{equation}
The total Hamiltonian $H$ can take a form identical to the Hamiltonian
in the case of normal metal leads by applying the following
transformation to the electron field operators separately in the left
and right leads respectively
\begin{align}
\psi_{R/L}(x) \rightarrow\ &e^{\pm i/\vf \cdot \int_{-L/2}^x \!\ud x'
\vec{M}(x')\cdot \vec \sigma} \psi_{R/L}(x), &x&<-{\textstyle
\frac{L}{2}} \nonumber \\  
\psi_{R/L}(x) \rightarrow\ &e^{\pm i/\vf \cdot \int_{L/2}^x \! \ud x'
\vec{M}(x')\cdot \vec \sigma} \psi_{R/L}(x)\mbox{,} &x&>{\textstyle
\frac{L}{2}}
\end{align}
This transformation leaves ${\mathcal H}_{\mathrm{LL}}$ invariant and
${\mathcal H}_{\mathrm{FM}}$ transforms into ${\mathcal H}_0$.

We apply the usual bosonization procedure to study this
model\cite{Haldane1,Haldane2}. The four electron modes are associated
to four bosonic modes described by the fields $\theta_{a\alpha}$ and
their duals $\varphi_{a\alpha}$ via the bosonization transformation
\begin{equation}\label{bosonize}
\psi_{R/L a\alpha} = \frac{1}{\sqrt{2\pi \Lambda }}\,e^{i (\varphi_{a
\alpha}\pm \theta_{a\alpha})}\ ,
\end{equation}
where $\Lambda$ is a short-distance cutoff. 
It is convenient to consider the following linear combinations of the
fields\cite{kbf}: $\theta_{i,c/s}=\frac{1}{\sqrt{2}} \left(
\theta_{i,\uparrow} \pm \theta_{i,\downarrow} \right)$ and
$\theta_{\pm,\mu}=\frac{1}{\sqrt{2}} \left( \theta_{1,\mu} \pm
\theta_{2,\mu} \right)$, with $i=1,2$ and $\mu=c,s$.  This allows us
to define the new fields $\theta_1=\theta_{+c}$, which corresponds to
the total charge mode and is the only interacting mode, and
$\theta_2=\theta_{+s}$, $\theta_3=\theta_{-c}$ and
$\theta_4=\theta_{-s}$; with similar transformations for the $\varphi$
fields.  In terms of these new fields the Luttinger liquid Hamiltonian
density (\ref{H}) is diagonalized
\begin{align}\label{HLL}
{\mathcal H}_{\mathrm{LL}}=& \frac{v}{2\pi} \left[g \left(\partial_x
\varphi_1 \right)^2 + \frac{1}{g} \left(\partial_x \theta_1 \right)^2
\right] \nonumber \\& +\sum_{i=2}^4 \frac{\vf}{2\pi} \left[
\left(\partial_x \varphi_i \right)^2 + \left(\partial_x \theta_i
\right)^2 \right]\ ,
\end{align}
where $\vf$ is the Fermi velocity, $v$ is the renormalized velocity
due to the interactions and $g$ is the Luttinger liquid parameter. In
the inhomogeneous model $g$ and $v$ are functions of the position:
$g=1$ and $v=\vf$ in the leads and
$g=\sqrt{\frac{\vf}{\vf+8\lambda/ \pi }}<1$ and $v=\vf/g$ in the
nanotube. 

The contacts between the leads and the nanotube are modeled by weak
backscattering at the contact points, the corresponding Hamiltonian
density has the form
\begin{widetext}
\begin{align}\label{Hbs} 
{\mathcal H}_{\mathrm{bs}} &= \sum_{m=1}^2 \sum_{a,b=1}^2
\sum_{\alpha,\beta=\uparrow,\downarrow} \delta(x-x_m) \left[ \tilde
u_m^{a b}\, \psi_{La\alpha}^\dag \,\psi_{Rb\alpha}^{\vphantom{\dag}} +
\ \tilde v_m^{a b}\, \psi_{La\alpha}^\dag \,\vec{M}_m \cdot \vec
\sigma_{\alpha \beta} \,\psi_{Rb\beta}^{\vphantom{\dag}}\ +\
{\mathrm{h.c.}}  \right] 
\\ &
= \sum_{m,a,b=1}^2 \sum_{\alpha=\pm 1} \delta(x-x_m)
 \left\{ \left( u_m^{a b} +\alpha v_m^{a b} M_m^z \right)
e^{ i \left[\theta_1+\alpha\theta_2+(-1)^{a+1} \delta_{a b} \left(
\theta_3+\alpha\theta_4 \right)+ (-1)^{a+1} \left( 1-\delta_{a b}
\right) \left( \varphi_3+\alpha \varphi_4 \right)\right] }
\right. \nonumber \\& \ \ 
 +  v_m^{a b} \left( M_m^x+i\alpha M_m^y
\right) e^{ i \left[\theta_1 + (-1)^{a+1} \delta_{a b} \theta_3 +
(-1)^{a+1} \left( 1 - \delta_{a b} \right) \alpha \theta_4 + \alpha
\varphi_2 + (-1)^{a+1} \left( 1-\delta_{a b} \right) \varphi_3+
(-1)^{a+1} \delta_{a b} \alpha \varphi_4 \right] }
%
+\, {\mathrm{h.c.}} \left. \vphantom{e^{ i \theta_1^2}}
\right\}  \ , \nonumber
\end{align}
\end{widetext}
where $m$ labels the position of the contacts: $x_{1/2}=\mp L/2$ ($L$
is the nanotube length) and $u_m^{a b}$ and $v_m^{a b}$ are constants
proportional to the strength of the backscattering, $u_m^{ab} = \tilde
u_m^{ab}/(2\pi\Lambda)$ and the same for $v_m^{a b}$.  The
backscattering terms are restricted by symmetry according to charge
conservation and spin rotational symmetry around the axis of
magnetization of the ferromagnetic contact.  We consider only terms of
the form $\psi_R^\dag \psi_L^{\vphantom{\dag}}$ because these are the
most relevant in the renormalization group sense (the scaling
dimension in real space of these terms is $\Delta=(g+3)/4$, while the
scaling dimension of terms of the form $\psi_{R/L}^\dag
\psi_{R/L}^{\vphantom{\dag}}$ is $\Delta=1$). Hence if all scattering
terms are weak, these terms will dominate. It is straightforward to
extend the present treatment to include the neglected interactions,
though we do not attempt this here.

The effect of the magnetization appears only on the backscattering
term.  In the case of near-perfect contacts to the electrodes, we can
treat the backscattering Hamiltonian $H_{\mathrm{bs}}$ as a
perturbation to the Hamiltonian $H=H_{\mathrm{LL}}$. This
procedure is described in section \ref{greens} in the context of the
Keldysh formalism that we  use in order to account for the effects
of the finite bias voltage.

The gate voltage introduces a term in the Hamiltonian density
proportional to
$\rho \, V_{\!g} = 
2/\pi \, V_{\!g}\,\partial_x \theta_1$, where $\rho$ is the electron
density. The constant of proportionality relates the voltage applied
at the gate with the voltage felt by the nanotube and therefore
depends on the sample.  The Hamiltonian density ${\mathcal
H}={\mathcal H}_{\mathrm{LL}}+{\mathcal H}_{V_{\!g}}$ becomes
${\mathcal H}={\mathcal H}_{\mathrm{LL}}$, after applying the following
transformation to the $\theta_1$ field:
$\theta_1\longrightarrow\theta_1- V_{\!g} \,x\ $, where we absorbed a
constant of proportionality into the definition of $V_{\!g}$ for
simplicity. This transformation needs to be applied to the total
Hamiltonian, including the backscattering term $H_{\mathrm{bs}}$, which
means that the gate voltage after this transformation will only
contribute to the perturbation Hamiltonian.

The effect of the external magnetic field is introduced via a Zeeman
coupling term in the Hamiltonian
\begin{equation}
H_h= - \vec h \cdot \int \dx \left( \psi^\dag_{R a \alpha}
\vec \sigma_{\alpha\beta} \psi^{\vphantom\dag}_{R a \beta} +
\psi^\dag_{L a \alpha} \vec \sigma_{\alpha\beta}
\psi^{\vphantom\dag}_{L a \beta} \right) \ .
\end{equation}
The contribution of the magnetic field can be transfered to the
perturbation Hamiltonian $H_{\mathrm{bs}}$ using a similar procedure to
the one described above for the gate voltage. Taking the z-direction
as the direction of the magnetic field, i.e.\ $\vec h =h \hat z$, the
Zeeman Hamiltonian density becomes ${\mathcal H}_h=- (h/\pi)
\partial_x \theta_2$ and applying $\theta_2 \rightarrow \theta_2 + B
x$ (with $B =2 h/\vf$), to the Hamiltonian $H=H_{LL}+H_{h}$, it
transforms as $H\rightarrow H_{LL}$. The results for non-zero magnetic
field are presented in appendix \ref{ap:magn}.

\section{\label{greens} The non-equilibrium transport problem}

Due to the finite bias voltage the distribution in this system is not
in thermal equilibrium.  This non-equilibrium situation is studied
using the Keldysh formalism (for a review, see
Ref.~\onlinecite{keldysh}).  To define a non-equilibrium initial
state, we assume that until some initial time $t_0$, the system has
reached quasi-equilibrium {\textsl{ in the absence of impurity
scattering}} $u_m^{ab}=v_m^{ab}=0$.  Without impurity scattering, the
total number of right- and left-moving carriers, $N_R$, $N_L$, are
separately conserved, so that a partial equilibrium can be established
with well-defined separate chemical potentials for the right and left
movers.  Hence, the system can be described, up to this time, by a
thermal distribution governed by the grand canonical Hamiltonian
\begin{equation}\label{truthfull}
  H_V = H_{\mathrm{LL}}- \mu_1 N_R- \mu_2 N_L \ , 
\end{equation}
where the chemical potentials in each lead are taken to be
$\mu_{1/2}=\mp V/2$ and $N_{R/L} = \int\dx n_{R/L}$.  The right
and left moving particle densities are given in the bosonization
procedure by $n_{R/L}= \frac{1}{2\pi}\sum_{a\alpha} \partial_x
\left(\pm  \varphi_{a\alpha}+ \theta_{a\alpha} \right)$. Then
\begin{equation}\label{HV}
H_V= H_{\mathrm{LL}}-\int\!\! \ud x \, \frac{2 V}{\pi}\, \partial_x
\varphi_1 \ .
\end{equation}
We emphasize that the appearance of the voltage $V$ in $H_V$ does not
represent a physical force on the electrons, but rather parameterizes
their non-equilibrium distribution.

After the initial time $t_0$ the evolution of the system is governed
by a different Hamiltonian $H$, which as deduced in section
\ref{model} is $H=H_{\mathrm{LL}}+H_{\mathrm{bs}}$, with
$H_{\mathrm{LL}}$ given in (\ref{HLL}) and $H_{\mathrm{bs}}$ in
(\ref{Hbs}).  We expect on physical grounds that introducing localized
scattering at the ends of the wire or nanotube reduces the current,
but cannot affect the non-equilibrium distribution in the reservoirs.
Hence, we believe that the prescription of defining the voltage $V$
from the initial distribution as is done using (\ref{truthfull}) gives
a faithful description of ideal leads.  According to this
prescription, a physical observable, represented by an operator
$\hat{\mathcal O}$, is then calculated from
\begin{equation}\label{eq:expect}
  \langle {\mathcal O}\rangle = \frac{1}{Z} {\mathrm{Tr}} \left( e^{-\beta
      H_V} e^{i H t} \hat{\mathcal O} e^{-i H t}\right) \ ,
\end{equation}
where
\begin{equation}\label{eq:pf}
 Z= {\mathrm{Tr}} \left( e^{-\beta H_V}\right) \ . 
\end{equation}
The difficulty in evaluating such an expectation value is that, unlike 
in a conventional equilibrium calculation, the Hamiltonian $H_V$
governing the initial distribution is different from $H$, which
governs the time evolution.  Thus such an expectation value cannot be
evaluated by equilibrium Green's function techniques.

Instead, we take advantage of the special property of $H_V$ that the
voltage couples only to $N_{R/L}$, which are decoupled ``zero mode''
degrees of freedom.  This technique has been applied a number of times
before to related problems\cite{KaneFisher,entangNT,andreev}, but to
our knowledge the details of its derivation have never been published.
For completeness, pedagogical value, and to highlight the physical
assumptions of the method, we include a thorough derivation in
appendix \ref{ap:derivation}.  The correction to $\langle {\mathcal
O}\rangle$ due to the backscattering is given by
\begin{equation}\label{eq:expect4} 
  \langle \delta {\mathcal O}\rangle = \frac{1}{Z_{\mathrm{LL}}}
    {\mathrm{Tr}} \left( e^{-\beta H_{\mathrm{LL}}}
    S^\dagger(t)\hat{\mathcal O} S(t)\right)\ ,
\end{equation}
where 
\begin{equation}\label{eq:evol}
  S(t) = {\mathrm{T}}\,\exp \left[ -i \int_0^t \!
    \ud t'\, H_I(t') \right]
\end{equation}
is the evolution operator for a system with the time-dependent
Hamiltonian $H_I(t')$.  Here $H_I(t)$ is the Hamiltonian in the frame
co-moving with the ideal current, defined by
\begin{equation}\label{eq:HI}
  H_I(t) = e^{i t \hat{V}} H e^{-i t \hat{V}} = H_{\mathrm{LL}} + \left[
  H_{\mathrm{bs}} \right]_{\theta_1 \rightarrow \theta_1 + V t} \ ,
\end{equation}   
with $\hat{V} = \frac{V}{2} (N_R-N_L)$.  Note that (because
$[\hat{V},H_{\mathrm{LL}}]=0$) all the time dependence in $H_I(t)$ is in
the backscattering term, and is hence easy to handle when working
perturbatively in $H_{\mathrm{bs}}$.

Eqs.~(\ref{eq:expect4}-\ref{eq:HI}) provide a reformulation of the
transport problem which is particularly convenient for a perturbative
treatment of the backscattering.  Note that -- because the voltage $V$
appears only within $H_{\mathrm{bs}}$ -- a direct expansion of
Eq.~(\ref{eq:expect4}) in $H_{\mathrm{bs}}$ will involve
{\textsl{equilibrium}} real-time propagators calculated with respect
to $H_{\mathrm{LL}}$.  We develop this perturbation theory using the
Keldysh path integral formulation.  This involves the usual
Trotterization of the two evolution operators $S^\dagger,
S^{\vphantom\dagger}$ in Eq.~(\ref{eq:expect4}) using coherent-state
fields denoted $\theta^+,\varphi^+$ for $S$ (``forward branch'') and
$\theta^-,\varphi^-$ (``backward branch'') for $S^\dagger$.  Further
noting that $H_{\mathrm{LL}}$ is quadratic and $H_{\mathrm{bs}}$ acts
only at the ends of the nanotube/wire, the fields away from $x=\pm
L/2$ can be integrated out to obtain the Keldysh integral
\begin{equation}
   \langle \delta {\mathcal O}\rangle = \int\! {\mathcal
     D}[\theta^\pm(t)\varphi^\pm(t)] {\mathcal O}_K e^{i S_0 - i
     \int\! \ud t H_{\mathrm{pert}}(t) } \ ,
\end{equation}
with
\begin{equation}
  H_{\mathrm{pert}}= H_{\mathrm{bs}}[\varphi_{i}^+,\theta_{i}^+
  +\delta_{i1} V t] - H_{\mathrm{bs}}
  [\varphi^-_{i},\theta^-_{i}+\delta_{i1} V t] \ .
\end{equation}
Here ${\mathcal O}_K$ is an appropriate Keldysh representation of the
operator ${\mathcal O}$, which can be chosen as usual from fields
lying on either the forward or backward moving branch, or any linear
combination thereof -- see below for convenient choices.  The
quadratic action $S_0$ is a functional of
$\theta^\pm(t),\varphi^\pm(t)$, which can be determined from the fact
that it must reproduce the {\textsl{equilibrium}} correlation and response
functions for these fields.  Indeed, we do not require an explicit
expression for $S_0$, but instead give the response and correlation
functions, defined by
\begin{align}\label{Gth}
C^\theta(x,t;x',t')&= \langle \theta (x,t) \theta(x',t') \rangle =
\tfrac{1}{2} \langle \{ \hat\theta (x,t), \hat\theta(x',t') \} \rangle
\ ,\nonumber
\\ 
R^\theta(x,t;x',t')&= \langle \theta (x,t)\tilde\theta(x',t') \rangle
= \nonumber 
\\ 
&=-i \Theta(t-t') \langle [ \hat\theta (x,t), \hat\theta(x',t') ]
\rangle \ ,
\end{align}
where we have applied the standard Keldysh rotation to the fields
$\theta^{\pm}=\theta \pm \frac{i}{2} \,\tilde\theta$. By construction
$\langle \tilde\theta(x,t)\tilde\theta(x',t')\rangle=0$. 
The Green's functions involving the $\varphi$ fields are defined in a
similar way,   replacing $\theta$ by $\varphi$ in the above
equations. 
There are also Green's functions that involve both $\theta$ and
$\varphi$, these are defined by 
\begin{align}\label{Gthph}
C^{\theta\varphi}(x,t;x',t')&= \langle \theta (x,t) \varphi(x',t')
\rangle = \tfrac{1}{2} \langle \{ \hat\theta (x,t), \hat\varphi(x',t')
\} \rangle\nonumber 
\\ 
R^{\theta\varphi}(x,t;x',t')&= \langle \theta
(x,t)\tilde\varphi(x',t') \rangle = \nonumber 
\\ 
&=-i \Theta(t-t') \langle [ \hat\theta (x,t), \hat\varphi(x',t') ]
\rangle \ ,
\end{align} 
and similar definitions for $C^{\varphi\theta}$ and
$R^{\varphi\theta}$. Again, by construction $\langle \tilde
\theta (x,t) \tilde \varphi(x',t') \rangle=\langle \tilde
\varphi (x,t) \tilde \theta(x',t') \rangle=0$.

Using the above procedure we obtain (up to additive constants
that will not contribute to the final result) the Green's functions
for the $\theta_1$ fields:
\begin{align}\label{thetagreens}
R^{\theta I}_{11}(t) &=-\frac{\pi}{2} (1-\alpha) \!\left[ \Theta(t)
+\frac{1+\alpha}{\alpha} \sum_{k \geq 1} \alpha^{2k}\, \Theta (t-2 k
t_v) \right] \nonumber 
\\ 
R^{\theta I}_{12}(t) &=-\frac{\pi}{2}
(1-\alpha^2) \sum_{k \geq 0} \alpha^{2k}\, \Theta (t- (2k+1) t_v) \ ,
\nonumber 
\\ 
C^{\theta I}_{11}(t) &= -\frac{1-\alpha}{4} \left[ \log
t^2+ \phantom{\sum_{k \geq 1}} \right. \nonumber \\ & \qquad\qquad
+\left.  \frac{1+\alpha}{\alpha} \sum_{k \geq 1} \alpha^{2k}\, \log
\left|t^2-(2 k t_v)^2 \right| \right] \ , \nonumber 
\\ 
C^{\theta I}_{12}(t) &= -\frac{1-\alpha^2}{4} \sum_{k \geq 0}
\alpha^{2k}\, \log\left|t^2- [ (2k+1) t_v ]^2 \right| \ ,
\end{align}
where the subscripts label the position of the contacts $x_{1,2}$
(e.g.\ $C_{ab}(t)=C(x_a,t;x_b,0)$) and 
\begin{equation}
\alpha=\frac{1-g}{1+g} \qquad {\mathrm{and}} \qquad t_v=\frac{L}{v} \ .
\end{equation}
We also need the Green's functions for the non-interacting modes
$\theta_{2,3,4}$, $R^F$ and $C^F$.  These are obtained from
(\ref{thetagreens}) by taking $\alpha=0$ and replacing $t_v$ by
$t_F=L/\vf$,
\begin{align}\label{freegreens}
R^{F}_{11}(t) &=-\frac{\pi}{2}  \Theta(t) \ , \nonumber 
\\ 
R^{F}_{12}(t) &=-\frac{\pi}{2} \Theta (t- t_F) \ , \nonumber 
\\ 
C^{F}_{11}(t) &= -\frac{1}{4} \log t^2 \ , \nonumber
\\
C^{F}_{12}(t) &= -\frac{1}{4}  \log \left|t^2- t_F^2 \right| \ .
\end{align}
The Green's functions for the $\varphi$ fields can be obtained
from those for the $\theta$ fields given in (\ref{thetagreens}), by
replacing $g$ by $1/g$, i.e.\ by   replacing $\alpha$ by
$-\alpha$. On the other hand, the only $\varphi$ Green's functions
that contribute to the transport properties studied in the following
sessions are those that correspond to the non-interacting modes
$\varphi_{2,3,4}$, and therefore they are identical to the functions
given in (\ref{freegreens}).

In order to compute the spin transport properties in section
\ref{spint}, we also need the following functions for the $\theta_2$
and $\varphi_2$ fields
\begin{align}\label{thfi}
R^{\theta_2 \varphi_2}(x,t) &= {\mathrm{sign}} (x)\, \frac{\pi}{2} \
\Theta(t) \ \Theta(|x|-\vf t) \ ,\nonumber \\ C^{\theta_2
\varphi_2}(x,t) &=- \frac{1}{4} \ \log \left|\frac{\vf t-x}{\vf t+x}
\right| \ ,
\end{align}
$C^{\varphi_2 \theta_2}(x,t)=C^{\theta_2 \varphi_2}(x,t)$ and
$R^{\varphi_2 \theta_2}(x,t)=R^{\theta_2 \varphi_2}(x,t)$.

\section{\label{charget} The differential conductance}

In this section we study the charge transport properties of 1D
electron systems and how these are affected by the magnetization of
the leads and, more importantly, the presence of the strong Coulomb
interactions.

We use the procedure described in sections \ref{model} and
\ref{greens} to calculate the differential conductance for these
systems. This is obtained from the expectation value of the current
in  a nanotube, i.e.\ a four mode 1D electron system with the Hamiltonian
given in (\ref{HLL}), as
\begin{equation}\label{Ix}
I= \sum_{a\alpha} \langle\psi_{Ra\alpha}^\dag
\psi_{Ra\alpha}^{\vphantom{\dag}} -\psi_{La\alpha}^\dag
\psi_{La\alpha}^{\vphantom{\dag}} \rangle
= \frac{2}{\pi} \langle \partial_t \theta_1 \rangle \ . 
\end{equation}

After a lengthy but straightforward calculation we obtain that the
differential conductance $G=\partial I/\partial V$ to second order in
perturbation theory is given by 
\begin{equation}\label{g2}
G = \frac{2}{\pi}\! \left\{\! 1 +\! \sum_{m}
U_m \int\!\!\ud t\, t\, e^{{\bm
C}_{1m}(t)} \sin\!\left[ \frac{1}{2} {\bm
R}_{1m}(t)\right]\cos(V t) \!\right\}
\end{equation}
with 
\begin{align}\label{uu}
U_1=& \sum_{m a b} \left[ \left(u^{a b}_m\right)^2 +
\left(v^{a b}_m\right)^2 \,\vec{M}_m^2 \right] \ , 
\\ U_2=& 2 \cos (\vg L) \sum_{a b} \left[ u^{a b}_1\, u^{a b}_2 +
v^{a b}_1\, v^{a b}_2 \, \vec{M}_1 \cdot \vec{M}_2
\right] \ , \nonumber
\end{align}
and 
\begin{equation}
{\bm C}_{a b}(t)=C^{\theta I}_{a b}(t) + 3 C^F_{a b}(t) \ ,
\end{equation}
and similarly for ${\bm R}_{a b}(t)$. For a quantum wire (i.e.\ a
single-channel electron system) these are replaced by ${\bm C}_{a
b}(t)=2 \left( C^{\theta I}_{a b}(t) + C^F_{a b}(t) \right)$ and the
global normalization is divided by a factor of two (since the quantum
wire has two modes instead of four).

Eqs.~(\ref{g2}) and (\ref{uu}) are valid for zero external magnetic
field, which is the case considered in this section, the equations for
non-zero magnetic field are presented in appendix \ref{ap:magn}.

\begin{figure*}
\begin{center}
\fig{.95\textwidth}{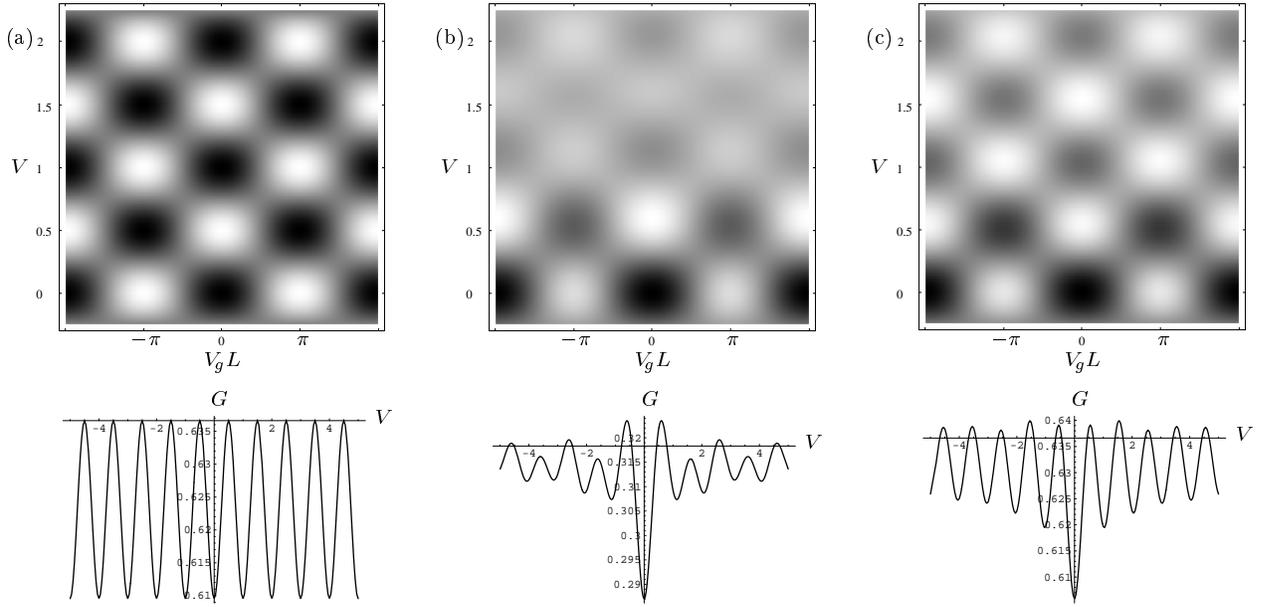}
\end{center}
\caption{Calculated conductance for identical contacts and $\vec
M_1=\vec M_2$ for (a) a free electron model, i.e.\ a nanotube with $g=1$;
(b) a quantum wire, i.e.\ a LL with two modes: spin and charge, with
$g=0.5$; and (c) a nanotube with $g=0.25$, as a function of bias $V$
and gate $\vg$ voltages (top) and as a function of the bias voltage
$V$ at constant gate voltage $\vg=0$ (bottom). As can be seen, the
effect of the interactions is quite appreciable, in particular with
the dependence in bias voltage at $\vg=0$. The ``period'' of these
oscillations is given by $2\pi/t_F$ ($t_F=2\pi$ in these figures), in
agreement with Ref.~\onlinecite{Liang}, but in (b) and (c) there is
another quasi-periodic component in these oscillations, with period
given by $2\pi/t_v$, the presence of these two time scales, $t_F$ and
$t_v$, is a direct result of spin charge separation.
\label{fig:G}}
\end{figure*}

Equation~(\ref{g2}) can  be easily generalized to arbitrary order in
perturbation theory, but the time integrals need to be computed
numerically. We present the calculated conductance to second order for
three different physical models in Fig.~\ref{fig:G}. This models
correspond to (a) a nano\-tube with non-interacting electrons, i.e.\
taking $g=1$, (b) a quantum wire with $g=0.5$ and (c) a nanotube with
$g=0.25$, which is a physically relevant value for single-walled
carbon nanotubes\cite{kbf,Bockrath,Yao1,Yao2,Dekker}.  The effect of
the interactions is visible in the dependence of the conductance with
bias voltage, at constant gate voltage.

The conductance is a quasi-periodic function of the bias voltage. At
$\vg =0$, for the non-interacting case, see Fig.~\ref{fig:G}.(a),
this dependence is a cosine function with period $2\pi/t_F$. For a
quantum wire, Fig.~\ref{fig:G}.(b), there are clearly two different
``periods'' in the oscillations, these are related to the two time
scales $t_F=L/\vf$ and $t_v=g t_F=L/v$. The existence of these two
different time scales is due to the two bosonic excitations in this
system: The spin excitation with velocity $\vf$ and the charge
excitation with velocity $v$, and is therefore an effect of spin
charge separation. The same effect appears in Fig.~\ref{fig:G}.(c),
but since for the nanotube there are three non-interacting modes with
velocity $\vf$ and only one mode, the total charge, with velocity $v$,
it is less visible than in the previous example. The most visible
effect of the interactions in the nanotube, is the enhancement of the
amplitude in the conductance around $V=0$. This effect is observable
in the experimental data presented in Ref.~\onlinecite{Liang}.

\begin{figure}[b]
\begin{center}
\vspace{.5cm}
\fig{.3\textwidth}{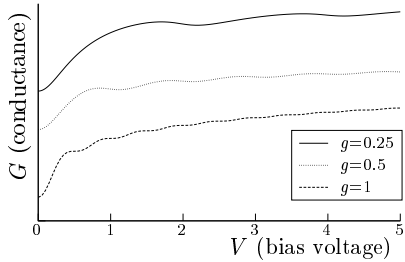}
\end{center}
\caption{\label{fig:pi2} 
At constant gate voltage: $V_g L=\pi/2$, the period of oscillations is
$\pi/g t_L$, i.e.\ depends strongly on the interaction strength. In
order for this effect to be clearly visible, we scaled and shifted the
functions differently, therefore the values of the $G$ axes are not
meaningful. }
\end{figure}

The calculated conductance at $\vg L =\pi/2$ as a function of the bias
voltage for a nanotube with different interaction
strengths corresponding to $g=0.25,\ 0.5$ and $1$, is presented in
Fig.~\ref{fig:pi2}. It can be seen using Eqs.~(\ref{g2}) and
(\ref{uu}) that the conductance for this value of the gate voltage
only depends on the Green's functions ${\bm C}_{11}$ and ${\bm
R}_{11}$, which depends only on $t_v$. As a result, we can clearly see
in Fig.~\ref{fig:pi2} that the period of the oscillations is
$\pi/t_v$, and therefore depends strongly on the interaction
strength. The amplitude of this oscillations is very small except for
the first oscillation, which is enough to identify this effect.

As for the dependence with the gate voltage, the conductance is a
periodic function, which is modulated by $\vec M_1 \cdot \vec M_2$,
and this is the main effect of the magnetization in the leads on the
conductance. In particular if $\sum u^{ab}_1 u^{ab}_2 \leq \sum
v^{ab}_1 v^{ab}_2$, there is an angle between the two magnetizations
for which $U_2$ vanishes for any value of the gate voltage, in this
case the conductance is given also in Fig.~\ref{fig:pi2}.

\section{\label{spint} Spin transport}

In this section we study spin transport properties, i.e.\ the spin
density in the nanotube and the spin current generated by the
magnetization in the leads.

The spin density expectation value in the nanotube, calculated using
bosonization and the Keldysh perturbation formalism as described in
sections \ref{model} and \ref{greens}, is given by 
\begin{equation}
\vec S = \frac{1}{2}\sum_{a\alpha\beta} \langle \psi_{Ra\alpha}^\dag
\, \vec \sigma_{\alpha \beta} \, \psi_{Ra\beta}^{\vphantom{\dag}} +
\psi_{La\alpha}^\dag\, \vec \sigma_{\alpha \beta} \,
\psi_{La\beta}^{\vphantom{\dag}} \rangle \ .
\end{equation}
For zero magnetic field, it is technically simplest to calculate
$S^{z}$ from the bosonized form
\begin{equation}\label{Szbos}
S^{z}= \frac{1}{\pi} \langle \partial_{x}\theta_{2} \rangle
\end{equation}
and then obtain the other two components by rotational invariance. 
For non-zero magnetic field the calculation as well as the final
results are much more involved, therefore and for the sake of clarity
we only present the results in appendix \ref{ap:magn}.

The result is 
\begin{align} 
 \vec S &= \, - \frac{1}{\vf} \left( \vec u_1 \int\!\!\ud t\, e^{{\bm
C}_{11}(t)} \sin\!\left[ \frac{1}{2} {\bm R}_{11}(t)\right]\sin(V t)
\right. \nonumber \\ & 
+ \sin (\vg L) \left\{\vec u_2 \int\!\!\ud t\, e^{{\bm C}_{12}(t)}
\sin\!\left[ \frac{1}{2} {\bm R}_{12}(t)\right]\sin(V t)
\right. \nonumber \\ & \left. \left. 
+\vec u_3 \int\!\!\ud t\, e^{{\bm C}_{12}(t)} \sin\!\left[ \frac{1}{2}
{\bm R}_{12}(t)\right]\cos(V t) \right\} \right) \ ,
\end{align}
with
\begin{align}\label{us}
\vec u_1 &=\sum_{a b} \left( u^{a b}_1 v^{a b}_1 \vec{M}_1 -
u^{a b}_2 v^{a b}_2 \vec{M}_2 \right)  \ , \nonumber \\
\vec u_2 &=\sum_{a b} v^{a b}_1 v^{a b}_2 \vec{M}_1
\times \vec{M}_2 \ , \nonumber \\
\vec u_3 &=\sum_{a b} \left( u^{a b}_1
v^{a b}_2 \vec{M}_2 + u^{a b}_2 v^{a b}_1\,
\vec{M}_1 \right) \ .
\end{align}

\begin{figure*}
\begin{center}
\fig{.95\textwidth}{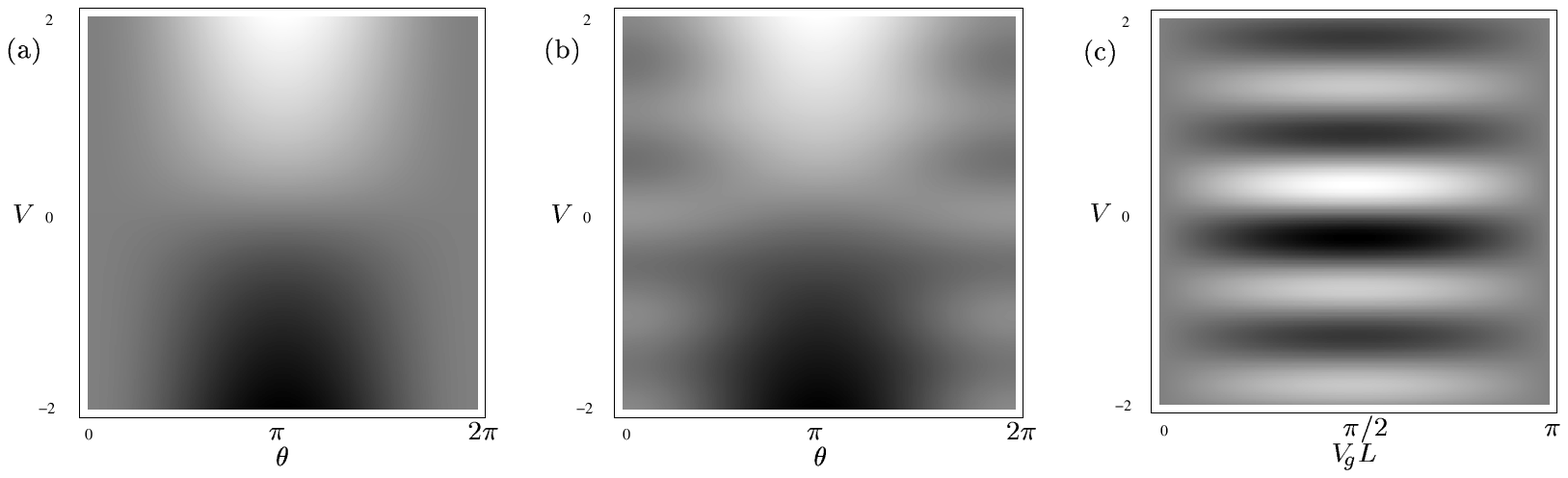}
\end{center}
\caption{ Calculated spin for a nanotube with $g=0.25$, the component
of the spin in the direction of the magnetization $\vec M_1$ as
function of the angle between the two magnetizations $\theta$ and the
bias voltage $V$, at constant gate voltage (a) $\vg =0$ and (b) $\vg
=\pi/2$; and (c) the component of the spin perpendicular to the plane
formed by the two magnetizations as a function of the gate and bias
voltages.
\label{fig:spin}}
\end{figure*}

Notice that the spin density does not depend on the position in the
nanotube, hence the total spin is   $L \vec S$.

The first term, proportional to $\vec u_1$, is the known
non-equilibrium spin accumulation
effect\cite{Aronov,Silsbee,Johnson,Prinz,Jay,Gijs}. It is maximum for
$\vec M_1=-\vec M_2$, when, in the case of identical contacts, the
other terms vanish. This term does not couple two backscatterers, is
independent of the gate voltage, and is an increasing function of the
bias voltage. It is depicted in Fig.~\ref{fig:spin}(a), since it is
the only term term that corresponds to the component of the spin in
the direction of $\vec M_1$ at $\vg=0$.  The second term corresponds
to the component of the spin perpendicular to the plane of the
magnetizations and is depicted in Fig.~\ref{fig:spin}(c), as function
of the bias and gate voltages.  The third term is the only one that
survives in equilibrium, i.e.\ at zero bias, it is due to the fact
that the backscattering strengths depend on the spins of the incoming
and outgoing electron relative to the direction of the magnetizations.
It is maximum for $\vec M_1=\vec M_2$, when again for identical
contacts the other terms vanish. This is depicted in
Fig.~\ref{fig:spin}(b), at $\theta=0,2\pi$. These terms that couple
the two backscatterers, and hence depend on the gate voltage, vary
with bias voltage in a manner approximately described by a sum of two
periodic functions, with ``periods'' given by $2\pi/t_F$ and
$2\pi/t_v$, as discussed in the previous section for the conductance.

The spin current 
\begin{equation}
\vec J_s= \frac{\vf}{2} \sum_{a\alpha\beta} \langle
\psi_{Ra\alpha}^\dag \, \vec \sigma_{\alpha \beta} \,
\psi_{Ra\beta}^{\vphantom{\dag}} - \psi_{La\alpha}^\dag\, \vec
\sigma_{\alpha \beta} \, \psi_{La\beta}^{\vphantom{\dag}} \rangle
\end{equation}
is as the spin density calculated from the $J^{z}_{s}$ component in
its bosonized form (again see result for non-zero magnetic field in
appendix \ref{ap:magn})
\begin{equation}\label{Jzbos}
J^{z}_s= \frac{1}{\pi} \langle \partial_t \theta_2 \rangle \ .
\end{equation}
It is not well-defined at the contact points because the
backscattering term in the Hamiltonian (\ref{Hbs}) does not conserve
spin, and therefore it has different expressions in the nanotube and
the leads.

The spin current in the left ($+$) and right ($-$) leads is given by
\begin{align}\label{Jleads}
\vec J_s = & \vec u_4 \int\!\!\ud t\, e^{{\bm
C}_{11}(t)} \sin\!\left[ \frac{1}{2} {\bm R}_{11}(t)\right]\sin(V t) 
\nonumber \\ & + \left[\pm \sin (\vg L)\ \vec u_2
+ \cos (\vg L)\, \vec u_3 \right] \nonumber\\ & \qquad \times
\int\!\!\ud t\, e^{{\bm C}_{12}(t)} \sin\!\left[ \frac{1}{2} {\bm
R}_{12}(t)\right]\sin(V t)  
\end{align}
and in the nanotube by:
\begin{align}
\vec J_s =& \vec u_4 \int\!\!\ud t\, e^{{\bm
C}_{11}(t)} \sin\!\left[ \frac{1}{2} {\bm R}_{11}(t)\right]\sin(V t) 
\\ & + \cos (\vg L)\,\left\{ \ \vec u_2 \int\!\!\ud t\, e^{{\bm
C}_{12}(t)} \sin\!\left[ \frac{1}{2} {\bm R}_{12}(t)\right]\cos(V t)
\right. \nonumber \\
& \quad\quad\qquad + \left. \vec u_3
\int\!\!\ud t\, e^{{\bm C}_{12}(t)} \sin\!\left[ \frac{1}{2} {\bm
R}_{12}(t)\right]\sin(V t) \right\} \ . \nonumber
\end{align}
with $\vec u_2$ and $\vec u_3$ defined in (\ref{us}) and 
\begin{equation}
\vec u_4= \sum_{m a b} u^{a b}_m v^{a b}_m \, \vec{M}_m \ .
\end{equation}

\begin{figure*}
\begin{center}
\fig{.95\textwidth}{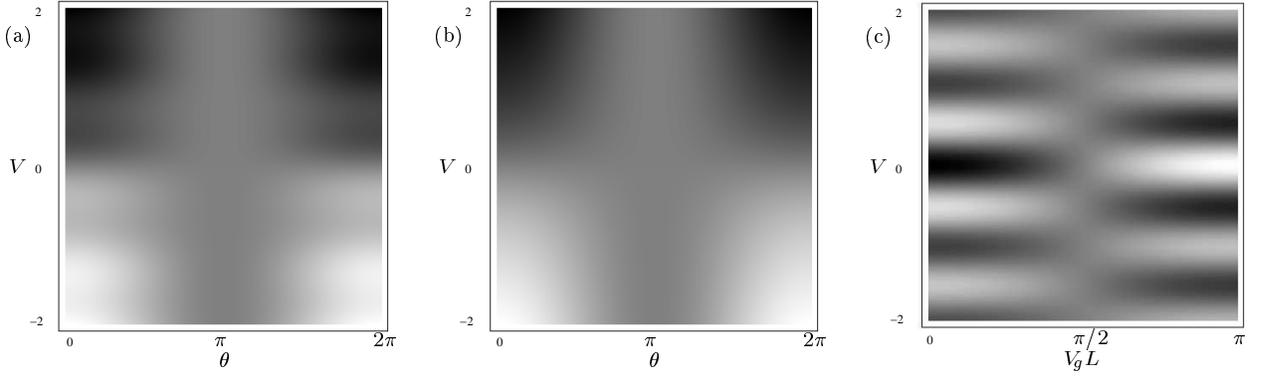}
\end{center}
\caption{ Calculated spin current for a nanotube with $g=0.25$: the
component of the spin current in the direction of the magnetization
$\vec M_1$ as function of the angle between the two magnetizations
$\theta$ and the bias voltage $V$, at constant gate voltage (a) $\vg
=0$ and (b) $\vg =\pi/2$; and (c) the component of the spin current
perpendicular to the plane formed by the two magnetizations as a
function of the gate and bias voltages (this figure corresponds to the
spin current in the nanotube, in the leads this component of the spin
current is identical to Fig.~\ref{fig:spin}.(c)).
\label{fig:spincurr}}
\end{figure*}

Similarly to the results for the spin discussed above, the first term,
which only involves one backscatterer and is independent of the gate
voltage, corresponds to the usual spin injection effect. It is an
increasing function of the gate voltage and is maximum for $\vec
M_1=\vec M_2$. This can be seen in Fig.~\ref{fig:spincurr}(b), since
it is the only term that does not vanish in the direction of $\vec
M_1$ at $\vg L=\pi/2$. At $\vg =0$, the terms proportional to $\vec
u_1$ and $\vec u_3$ contribute equally to the component of the current
in the direction of $\vec M_1$, the result for this case is presented
in Fig.~\ref{fig:spincurr}(a).  The second term, proportional to $\vec
M_1 \times \vec M_2$ corresponds to an exchange interaction between
the magnetizations of the leads, mediated by the nanotube. It has
opposite signs in the two leads and it is shown in
Fig.~\ref{fig:spincurr}(c).

\section{Conclusions}

We studied the charge and spin transport properties of 1D systems,
e.g.\ quantum wires and carbon nanotubes, focusing on the latter. We
considered the case of nearly perfect ohmic contact between the 1D
system and the electrodes and included the strong Coulomb interaction
via a Luttinger liquid model. We found important effects on the
transport properties of these systems that are due to the Coulomb
interactions. These appear in the dependence with bias voltage. In
particular, the conductance is enhanced at low bias voltage,
furthermore it is an oscillatory function where we can distinguish two
quasi-periodic components, with periods that are related to the two
velocities of the excitations of a Luttinger liquid, $v$ and $\vf$.
This effect is therefore a direct consequence of spin and charge
separation. It is clearly visible in single band quantum wires.  In
nanotubes, the amplitude of the higher period component is reduced by
the presence of three (as opposed to one) neutral modes.  Still, we
can find evidence of the two velocities $v$ and $\vf$ by comparing the
dependence of the conductance with bias voltage for two different gate
voltages ($\vg L=0$ and $\pi/2$).  It is perhaps worth noting that,
for the case of non-magnetic leads with symmetric contacts, the
conductance formula involves only two unknown parameters: the overall
amplitude of the backscattered current, and the Luttinger parameter
$g$, both of which can be simply estimated.  Nevertheless, a
non-trivial functional dependence upon bias voltage is predicted.
 
The spin and spin current have one component in the plane of the
magnetization that does not couple the two leads and is therefore
independent of the gate voltage. This term should be understood as
arising from incoherent spin injection at each contact.  It is a
monotonic function of the bias voltage, and corresponds to the known
spin accumulation (for the spin) and spin injection (for the spin
current) effects. The other components that couple the two leads, and
therefore depend on the gate voltage, are backscattering processes
occurring with coherence between the two contacts.  These oscillate
with the bias voltage, in a manner approximately described as a sum of
two periodic components, with periods related to the two velocities of
the excitation of the Luttinger liquid.  The amplitude of the
higher-period component is largest in a single-channel quantum wire,
and somewhat suppressed in nanotubes by the sub-band degeneracy.

\begin{acknowledgments}
C.S.P. was supported by FCT and FSE through grant PRAXIS/BD/18554/98.
L.B. was supported by NSF through grant DMR--9985255, and by the Sloan
and Packard foundations.
K.J.W. thanks the Deutsche Forschungsgemeinschaft for support through
Heisenberg grant Wi/1932 1-1.
\end{acknowledgments}

\appendix
\section{Derivation of Eq.~(\ref{eq:expect4})}\label{ap:derivation}
In this appendix, we derive Eq.~(\ref{eq:expect4}).  In particular, we
consider a large periodic system of size ${\textsf{L}}$, where ultimately
${\textsf{L}}\rightarrow \infty$.  We define the right/left-moving
combinations
\begin{equation}
  \phi_{i R/L} = \varphi_i \pm \theta_i \ .
\end{equation}
In the system of size ${\textsf{L}}$ we can decompose into finite
wavevector and ``zero mode'' components.  In particular, for the total 
charge fields, we define
\begin{eqnarray}
  \phi_{1R}(x) & = & \tilde\phi_{1R}(x) + \frac{2\pi N_R x}{\textsf{L}} +
  \Phi_R \ , \\
  \phi_{1L}(x) & = & \tilde\phi_{1L}(x) - \frac{2\pi N_L x}{\textsf{L}} +
  \Phi_L \ ,
\end{eqnarray}
where $\tilde\phi_{1R/L}(x)$ contains the non-zero momentum modes
of the $\phi_{1R/L}$ fields.  With these definitions, the zero-mode 
variables form two canonically conjugate pairs:
\begin{eqnarray}
  [N_R, \Phi_R]  & = & [N_L,\Phi_L]  =  i \ , \\ \nonumber
  [N_R, \Phi_L] &= & [N_L,\Phi_R] = [N_R,N_L]=[\Phi_R,\Phi_L]  =  0 \ .  
\end{eqnarray}
Moreover, $N_{R/L},\Phi_{R/L}$ commute with $\tilde\phi_{1R/L}(x)$
and all fields associated with channels $2,3,4$.  

Since the interactions which transform the system from a Fermi liquid
into a Luttinger liquid (Eq.~(\ref{H})) exist only for $|x|<L/2$, they 
do not affect the zero-mode terms in the Hamiltonian.  Hence one may
separate 
\begin{equation}
  H_V = \tilde{H}_{\mathrm{LL}} + \frac{\pi
    \vf}{\textsf{L}} \left( N_R^2 +
    N_L^2\right) - \frac{V}{2} \left( N_R-N_L \right) \ ,
\end{equation}
where $\tilde{H}_{\mathrm{LL}}$ is the Luttinger liquid Hamiltonian,
Eq.~(\ref{HLL}), with the zero-mode terms subtracted, i.e.\ with
$\varphi_1 \rightarrow \tilde\varphi_1$ and $\theta_1 \rightarrow
\tilde\theta_1$.  We then see, using the independence of the zero mode 
variables, that the unitary operator
\begin{equation}
  U_V = e^{i \frac{V{\textsf{L}}}{4\pi \vf}(\Phi_R - \Phi_L)} \ , 
\end{equation}
generates the following transformation $N_{R/L} \rightarrow N_{R/L}
  \pm V{\textsf{L}}/(4\pi \vf)$, hence
\begin{equation}
  e^{-\beta H_V} = e^{-{\mathcal C}}U_V e^{-\beta H_{\mathrm{LL}}}
  U_V^\dagger \ ,
\end{equation}
where ${\mathcal C}= \beta {\textsf{L}}/(8\pi \vf)$ is an unimportant
constant.  Inserting this into Eqs.~(\ref{eq:expect}-\ref{eq:pf}), one
obtains
\begin{equation}\label{eq:expect2}
  \langle {\mathcal O}\rangle = \frac{1}{Z_{\mathrm{LL}}} {\mathrm{Tr}}
      \left( e^{-\beta H_{\mathrm{LL}}} e^{i(H+ \hat{V})t}
      \left(U_V^\dagger \hat{\mathcal O}U_V\right)
      e^{-i(H+\hat{V})t}\right) \ ,
\end{equation}
with
\begin{equation}\label{eq:pf2}
 Z_{\mathrm{LL}}= {\mathrm{Tr}} \left( e^{-\beta H_{\mathrm{LL}}}\right) 
\end{equation}
and 
\begin{equation}\label{hatV}
  \hat{V} = \frac{V}{2} (N_R-N_L) \ .
\end{equation}
For the operators of interest,
\begin{eqnarray}
  U_V^\dagger \hat{I} U_V & = & I_0 + \hat{I} \ , \\
   U_V^\dagger \vec{S} U_V & = & \vec{S} \ , \\
   U_V^\dagger \vec{J}_s U_V & = & \vec{J}_s \ ,
\end{eqnarray}
where $I_0 = 4 (e^2/h) V = (2/\pi) V$ is the current which would flow
in an ideal nanotube in the absence of backscattering.  Defining
$\delta{\mathcal O} = {\mathcal O} - {\mathcal O}_0$, with ${\mathcal
O}_0 = I_0$ for ${\mathcal O}=\hat{I}$ and ${\mathcal O}_0=0$ for
${\mathcal O}=\vec{S},\vec{J}_s$, one has then
\begin{equation}\label{eq:expect3}
  \langle \delta {\mathcal O}\rangle = \frac{1}{Z_{\mathrm{LL}}}
    {\mathrm{Tr}} \left( e^{-\beta H_{\mathrm{LL}}} e^{i(H+
    \hat{V})t}\hat{\mathcal O}e^{-i(H+\hat{V})t}\right) \ .
\end{equation}
We then apply the formula 
\begin{equation}
  e^{-i t(H + \hat{V})} = e^{-i t \hat{V}}\ {\mathrm{T}}\, \exp \left[
    -i \int_0^t \!  \ud t'\, H_I(t') \right] \ ,
\end{equation}
to arrive at Eqs.~(\ref{eq:expect4}) ff.\ given in the main text. 

\section{Differential conductance, spin and spin current at non-zero
  magnetic field}\label{ap:magn} 

\begin{widetext}
The differential conductance including the magnetic field is still
given by Eqs.~(\ref{g2}) and (\ref{uu}) with only the following change
in $U_2$ 
%
\begin{align}
U_2=&2 \cos (\vg L) \sum_{ab} \left\{ u_1^{ab} u_2^{ab} 
+ v_1^{ab} v_2^{ab}\left[ \vec M_1\cdot \vec M_2
+ ( \vec M_1\cdot \hat h) ( \vec M_2\cdot \hat h) \left(\cos
B L -1 \right) 
\right]\right\}
\nonumber \\ &
+  \sin (B L) \sin (\vg L) \sum_{ab} \left[ u_1^{ab} v_2^{ab}
( \vec M_2 \cdot \hat h) 
%
%
+u_2^{ab} v_1^{ab} (M_1\cdot \hat h) \right] \ ,
\end{align}
where $\hat h =\vec h/h$. 

The total spin of the nanotube is 
\begin{align}
\vec S = &\frac{1}{\pi} \hat h B L - \frac{1}{\vf} \sum_{ab} \biggl(
\biggl\{ \left( u_1^{ab} v_1^{ab} \vec M_1 - u_2^{ab} v_2^{ab} \vec
M_2 \right) \tfrac{1}{B} \sin B L
%
+ \left( u_1^{ab} v_1^{ab} (\vec M_1 \cdot \hat h) \hat h - u_2^{ab}
  v_2^{ab} (\vec M_2 \cdot \hat h) 
\hat h \right)  \left( L - \tfrac{1}{B} \sin B L \right) 
\nonumber \\ & \qquad  \qquad \qquad
+ \left( u_1^{ab} v_1^{ab} (\vec M_1 \times \hat h) - u_2^{ab}
v_2^{ab} (\vec M_2 \times \hat h) \right) \tfrac{1}{B} \left( 1 - \cos
B L \right) \biggr\} \int_t e^{{\bm C}_{11}} \sin \left( \tfrac{1}{2}
{\bm R}_{11} \right) \sin (V t)
\nonumber \\ 
+ & \biggl\{ \left[ 
\left( u_1^{ab} v_2^{ab} \vec M_2 + u_2^{ab} v_1^{ab} \vec M_1 \right)
\tfrac{1}{B} \sin B L
%
+ \left( u_1^{ab} v_2^{ab} (\vec M_2 \cdot \hat h) \hat h + u_2^{ab}
v_1^{ab} (\vec M_1 \cdot \hat h) \hat h \right) \left( L \cos BL
-\tfrac{1}{B} \sin B L \right)
\right] \sin \vg L
\nonumber \\ & 
- u_1^{ab} u_2^{ab} \hat h L \sin B L \cos \vg L
- v_1^{ab} v_2^{ab} \left[ \left( (\vec M_1 \cdot \hat h) \vec M_2 +
  (\vec M_2 \cdot \hat h) \vec M_1 \right) \tfrac{1}{B} \left(1 - \cos
  B L \right)
\right. \nonumber \\ & \qquad  \qquad  \qquad  \left. 
+ (\vec M_1 \cdot \hat h)(\vec M_2 \cdot \hat h) \hat h \left( L \sin B
 L - \tfrac{2}{B} \left( 1 - \cos B L \right) \right) \right] \cos \vg L
\biggr\} \int_t e^{{\bm C}_{12}} \sin \left( \tfrac{1}{2} {\bm R}_{12}
\right)\cos (V t)
\nonumber \\ 
+ & \biggl\{ \left( u_1^{ab} v_2^{ab} (\vec M_2 \times \hat
h) - u_2^{ab} v_1^{ab} (\vec M_1 \times \hat h) \right) \tfrac{1}{B}
\left( 1 - \cos B L \right) \cos \vg L
%
+ v_1^{ab} v_2^{ab} \left[ \left( (\vec M_1\cdot\hat h) (\vec M_2
\times \hat h) - (\vec M_2 \cdot \hat h) (\vec M_1 \times \hat h)
\right)
\right. \nonumber \\ & \qquad \qquad \times \left.
\left( L-\tfrac{1}{B} \sin B L \right)
%
+ \vec M_1 \times \vec M_2 L \right] \sin \vg L \biggr\} \int_t e^{{\bm
C}_{12}} \sin \left( \tfrac{1}{2} {\bm R}_{12} \right)\sin (V t) \biggr) \ .
\end{align}

The spin current in the nanotube (Eq.~\ref{jsnt}) and the leads
(Eq.~\ref{jsl}) is 
\begin{align}\label{jsnt}
\vec J_s = \sum_{ab} & \biggl\{ u_1^{ab} v_1^{ab} \vec M_1 \cos B
(x-x_1) + u_2^{ab} v_2^{ab} \vec M_2 \cos B (x-x_2)
\nonumber \\ & 
+ u_1^{ab} v_1^{ab} (\vec M_1 \cdot \hat h) \hat h \left[ 1 - \cos B
(x-x_1) \right] + u_2^{ab} v_2^{ab} (\vec M_2 \cdot \hat h) \hat h
\left[ 1 - \cos B (x-x_2) \right]
\nonumber \\ & 
+ u_1^{ab} v_1^{ab} (\vec M_1 \times \hat h) \sin B (x-x_1)
%
- u_2^{ab} v_2^{ab} (\vec M_2 \times \hat h) \sin B (x-x_2) \biggr\}
\int_t e^{{\bm C}_{11}} \sin \left( \tfrac{1}{2} {\bm R}_{11} \right)
\sin (V t)
\nonumber \\
+ \sum_{ab} & \biggl\{ \Bigl[ u_1^{ab} v_2^{ab} (\vec M_2 \cdot \hat h)
    \hat h \left[ \cos B L - \cos B (x-x_1) \right] 
%
+ u_2^{ab} v_1^{ab} (\vec M_1 \cdot \hat h) \hat h \left[ \cos B L -
\cos B (x-x_2) \right]
\nonumber \\ & \qquad
+ u_1^{ab} v_2^{ab} \vec M_2 \cos B (x-x_1) + u_2^{ab} v_1^{ab} \vec
M_1 \cos B (x-x_2) \Bigl] \cos \vg L
\nonumber \\ & 
+ v_1^{ab} v_2^{ab} \Bigl[ (\vec M_1 \cdot \hat h)(\vec M_2 \cdot
\hat h) \hat h \left[ \sin BL - \sin B(x-x_1)+ \sin B(x-x_2) \right]
\nonumber \\ & \qquad 
+ (\vec M_1 \cdot \hat h) \vec M_2 \sin B(x-x_1) - (\vec M_2 \cdot
\hat h) \vec M_1 \sin B(x-x_2)
\nonumber \\ & \qquad 
+ u_1^{ab} u_2^{ab} \hat h \sin BL \Bigr] \sin \vg L \biggr\} \int_t
e^{{\bm C}_{12}} \sin \left( \tfrac{1}{2} {\bm R}_{12} \right) \sin (V
t)
\nonumber \\
+ \sum_{ab} & \biggl\{  \left[ u_1^{ab} v_2^{ab} (\vec M_2 \times \hat
h) \sin B (x-x_1) - u_2^{ab} v_1^{ab} (\vec M_1 \times \hat h) \sin B
(x-x_2) \right] \sin \vg L
\nonumber \\ & 
- v_1^{ab} v_2^{ab} \left[ (\vec M_1\cdot\hat h) (\vec M_2 \times
\hat h) \left[ 1-\cos B(x-x_1) \right] 
%
- (\vec M_2 \cdot \hat h) (\vec M_1 \times \hat h) \left[ 1-\cos
  B(x-x_2) \right]
\right. \nonumber \\ & \qquad \left.
+ \vec M_1\times \vec M_2 \right] \cos \vg L \biggr\} \int_t e^{{\bm
C}_{12}} \sin \left( \tfrac{1}{2} {\bm R}_{12} \right) \cos (V t) \ ,
\end{align}
\begin{align}\label{jsl}
\vec J_s = \sum_{ab} & \biggl\{ u_1^{ab} v_1^{ab} \vec M_1 \cos B
(x-x_1) + u_2^{ab} v_2^{ab} \vec M_2 \cos B (x-x_2)
\nonumber \\ & 
+ u_1^{ab} v_1^{ab} (\vec M_1 \cdot \hat h) \hat h \left[ 1 - \cos B
(x-x_1) \right] + u_2^{ab} v_2^{ab} (\vec M_2 \cdot \hat h) \hat h
\left[ 1 - \cos B (x-x_2) \right]
\nonumber \\ & 
\mp \left[ u_1^{ab} v_1^{ab} (\vec M_1 \times \hat h) \sin B (x-x_1) 
%
+ u_2^{ab} v_2^{ab} (\vec M_2 \times \hat h) \sin B (x-x_2)\right]
\biggr\} \int_t e^{{\bm C}_{11}} \sin \left( \tfrac{1}{2} {\bm R}_{11}
\right) \sin (V t)
\nonumber \\
+ \sum_{ab} & \biggl\{ \Bigl[ u_1^{ab} v_2^{ab} (\vec M_2 \cdot \hat h)
    \hat h \left[ \cos B L - \cos B (x-x_1) \right] 
%
+ u_2^{ab}
    v_1^{ab} (\vec M_1 \cdot \hat h) \hat h \left[ \cos B L - \cos B
    (x-x_2) \right]
\nonumber \\ & \qquad 
+ u_1^{ab} v_2^{ab} \vec M_2 \cos B (x-x_1) + u_2^{ab} v_1^{ab} \vec
M_1 \cos B (x-x_2) \Bigl] \cos \vg L
\nonumber \\
& + v_1^{ab} v_2^{ab} \Bigl[ (\vec M_1 \cdot \hat h)(\vec M_2 \cdot
\hat h) \hat h \left[ \sin BL - \sin B(x-x_1)+ \sin B(x-x_2) \right]
\nonumber \\ & \qquad 
+ (\vec M_1 \cdot \hat h) \vec M_2 \sin B(x-x_1) - (\vec M_2
\cdot \hat h) \vec M_1 \sin B(x-x_2) \Bigr] \sin \vg L
\nonumber \\
& \mp \left[ u_1^{ab} v_2^{ab} (\vec M_2 \times \hat h) \sin B (x-x_1)
+ u_2^{ab} v_1^{ab} (\vec M_1 \times \hat h) \sin B (x-x_2) \right]
\cos \vg L
\nonumber \\ &
\mp v_1^{ab} v_2^{ab} \left[ (\vec M_1\cdot\hat h) (\vec M_2 \times
\hat h) \left[ 1-\cos B(x-x_1) \right] 
%
- (\vec M_2 \cdot \hat h) (\vec
M_1 \times \hat h) \left[ 1-\cos B(x-x_2) \right]
\right. \nonumber \\ & \qquad \left.
+ \vec M_1\times \vec M_2 \right] \sin \vg L + u_1^{ab} u_2^{ab} \hat
h \sin BL \sin \vg L \biggr\} \int_t e^{{\bm C}_{12}} \sin \left(
\tfrac{1}{2} {\bm R}_{12} \right) \sin (V t) \ ,
\end{align}
where the $\mp$ sign correspond to lead 1,2 respectively.  

One can verify that the spin current and spin density are not
independent and are in fact related by the precessional equation of
motion,
\begin{equation}
  \label{eq:precess}
  \partial_t \vec{S} + \partial_x \vec{J}_s = - 2 \vec{h} \times \vec{S}.
\end{equation}
In the steady state, $\langle \partial_t \vec{S}\rangle =0$, so one
has
\begin{equation}
  \label{eq:precess2}
    \partial_x \langle \vec{J}_s(x) \rangle = - 2 \vec{h} \times
    \langle \vec{S}(x)\rangle.
\end{equation}
\end{widetext}


\end{document}